\documentclass[sigconf, nonacm]{acmart}

\usepackage{wrapfig}

\usepackage{float}
\usepackage{array}
\usepackage{booktabs}
\usepackage{svg}

\usepackage{graphicx}
\usepackage{subcaption}
\usepackage{wrapfig}

\usepackage[linesnumbered,ruled,vlined]{algorithm2e}
\SetKwComment{Comment}{/* }{ */}
\SetKw{Continue}{continue}
\SetKw{Break}{break}
\SetKw{Not}{not}

\newcommand{\low}{back\-ground}
\newcommand{\high}{time-sensitive}

\newcommand{\High}{Time-sensitive}

\newcommand{\RR}{SCHED\_RR}
\newcommand{\FIFO}{SCHED\_FIFO}

\newcommand{\runnabletree}{runnable tree}
\newcommand{\Runnabletree}{Runnable tree}

\newcommand{\myparagraph}[1]{%
  \par\addvspace{0.2em}%
  \noindent\emph{#1.}\hspace{0.1cm}%
}

\newenvironment{new-item}%
{\begin{list}{$\bullet$}{%
\topsep0pt \partopsep0pt \parsep0pt \itemsep0.1ex plus 0.1ex minus 0.1ex
\setlength{\leftmargin}{0.8\leftmargin}
\setlength{\labelsep}{0.8\labelsep}}}{\end{list}}

\newenvironment{new-enum1}%
{\begin{list}{\arabic{enum1}.}{\usecounter{enum1}%
\topsep0pt \partopsep0pt \parsep0pt \itemsep0.2ex plus 0.1ex minus 0.1ex
\setlength{\leftmargin}{0.8\leftmargin}
\setlength{\labelsep}{0.8\labelsep}}}{\end{list}}

\newenvironment{new-enum2}%
{\begin{list}{\alph{enum2}.)}{\usecounter{enum2}%

\setlength{\leftmargin}{0.8\leftmargin}
\setlength{\labelsep}{0.8\labelsep}}}{\end{list}}

\newenvironment{new-enum3}%
{\begin{list}{\roman{enum3}.}{\usecounter{enum3}%
\topsep0pt \partopsep0pt \parsep0pt \itemsep0.2ex plus 0.1ex minus 0.1ex
\setlength{\leftmargin}{0.8\leftmargin}
\setlength{\labelsep}{0.8\labelsep}}}{\end{list}}

\begin{document}

\title[Unfair by design: eBPF-based scheduling of mixed database workloads]{Unfair by design: eBPF-based scheduling\break of mixed database workloads}

\author{Carl-Elliott Bilodeau-Savaria}
\affiliation{%
  \institution{McGill University}
 \city{Montreal}
  \country{Canada}
}
\email{carl-elliott.bilodeau-savaria@mail.mcgill.ca}

\author{Jan Kristof Nidzwetzki}
\orcid{0000-0002-1825-0097}
\affiliation{%
  \institution{FernUniversität in Hagen}
  \streetaddress{}
  \city{Hagen}
  \country{Germany}
}
\email{jan-kristof.nidzwetzki@fernuni-hagen.de}

\author{Stefanie Scherzinger}
\affiliation{%
  \institution{University of Passau}
  \city{Passau}
  \country{Germany}
}
\email{stefanie.scherzinger@uni-passau.de}

\author{Bettina Kemme}
\affiliation{%
  \institution{McGill University}
 \city{Montreal}
  \country{Canada}
}
\email{bettina.kemme@mcgill.ca}

\begin{abstract}

Modern database systems increasingly co-schedule time-sensitive and background tasks. In such mixed workloads, background tasks should ideally utilize only spare CPU capacity without interfering with latency-critical requests. While some database-level solutions address this challenge, many database systems still rely on operating system (OS) schedulers, which, despite supporting priorities, do not reliably isolate high-priority tasks. Furthermore, they remain vulnerable to priority inversion, where preempted \low{} tasks can delay other work.
We present \textbf{UFS}, a selectively unfair scheduler implemented as an eBPF-based \texttt{sched\_ext} scheduler in the Linux kernel. UFS restricts background tasks to idle CPU capacity and preempts them immediately when time-sensitive tasks arrive. To address priority inversion, UFS incorporates application-level hints via eBPF maps, ensuring that background tasks are not unnecessarily delayed should \high{} tasks wait for them to release locks.
Our integration of UFS into PostgreSQL demonstrates that, under mixed workloads, UFS improves throughput for \high{} tasks by up to 2$\times$, while reducing tail latency by half, compared to existing scheduling options in Linux.

\end{abstract}

\maketitle

\section{Introduction}

Database management systems (DBMS) provide increasingly complex support for in-DBMS data analytics and data science processing~\cite{UDFTutorial,madlib,aida,InferDB,inDBinference,MLPredinDB}. In particular, user-defined functions (UDFs)~\cite{UDFBench,UDFTutorial} have become popular to implement not only pipelines of complex SQL queries and materialized view refreshes, but also to run ML inference~\cite{InferDB,inDBinference,MLPredinDB} or even ML learning tasks within the DBMS, close to the data~\cite{aida,madlib,Yunjia}.
These tasks are often long-running and CPU-bound, but might not require stringent response times. At the same time, DBMS must continue to ensure timely responses for interactive queries (often with bursty CPU usage patterns) and for certain CPU-bound tasks such as complex queries. 

Executing low- and high-priority database workloads concurrently is challenging. When long-running analytics significantly delay short requests, SLAs may be violated. The general solution is priority-based scheduling, which has been widely explored by both the database and the operating system (OS) communities. The scheduler manages tasks that are not time-critical as low-priority. They are executed when free CPU cycles are available, but should be preempted whenever high-priority tasks need the CPU.

Work on priority-based scheduling within the database system ~\cite{CareyJL89,McWherter05,polaris,sigmodcomp} often puts a particular focus on the priority inversion problem~\cite{57058}, where a low-priority task that holds a lock is preempted, causing a high-priority task waiting for this lock to be unduly delayed. The general solution is to avoid the preemption of low-priority tasks as long as they hold locks.

The OS community, in contrast, has developed a plethora of techniques where processes/threads with a higher priority get a larger share of resources. Today's Linux kernels feature \texttt{cgroups}, with a sophisticated mechanism for implementing complex hierarchical resource-control structures (e.g., CPU, I/O, and memory limits). 
The default scheduler in Linux, EEVDF, provides proportional sharing of the CPU and can be combined with \texttt{cgroups} to provide processes with resources proportional to the priority of their group. Furthermore, Linux supports dedicated scheduling algorithms for real-time tasks or to exploit idle CPU resources for background tasks. 

The current state-of-the-art in modern OS schedulers
motivated us to revisit 
OS-based scheduling for mixed database workloads. OS-based scheduling has advantages, %
as it can also consider non-DBMS processes or threads, e.g., kernel threads. Furthermore, high-priority tasks might not only be blocked when trying to access locks, but also when waiting for user input or I/O. In these situations, an OS scheduler can ensure that a low-priority task is granted the CPU. Thus, scheduling can make more holistic decisions. Finally, offering an OS solution is promising for existing database systems, since %
the integration of a DBMS-based scheduler is a significant engineering effort. Given all these advantages, an OS-based scheduler must nevertheless be able to handle the priority inversion problem, yet this only materializes in the DBMS. As such, some form of communication between DBMS and OS is required.

\textit{State-of-the-art OS schedulers.}
In a first step,
we explored the existing OS scheduling options for mixed database workloads, not even targeting the priority inversion problem yet. Our goal was to be completely \textit{unfair} upon request: \low\ tasks should only get the CPU when no \high\ tasks require computation. However, when there are different types of \high\ tasks, they should get a proportional share of the CPU. Unfortunately, and as also repeatedly observed by previous work~\cite{DBLP:conf/eurosys/LeverichK14,LoziWastedCoresPatchesNote, ZijlstraLKMLWastedCores2016}, the Linux scheduling options do not achieve the desired outcome.  With Linux’s proportional
scheduler EEVDF, even when CPU-bound background
tasks are given low priority, they negatively affect CPU-bursty high-priority tasks. On the other hand, Linux’s real-time scheduling algorithms disproportionally delay such CPU-bursty tasks in the presence of
CPU-bound tasks of the same priority.  %

\textit{Selectively Unfair Scheduler.} Our observations motivate us to design our own scheduler policy. In fact, recent Linux kernels allow new scheduling algorithms to be implemented through the \texttt{sched\_ext} interface as eBPF programs, which can be loaded at runtime. This allows for deployment and interaction without patching or rebuilding the Linux kernel, a very appealing opportunity for database-oriented scheduling.

We thus present the \emph{Unfair Scheduler} (UFS), a \texttt{sched\_ext}-based scheduler policy implementation for mixed database workloads. UFS is designed to be \emph{selectively unfair}: It grants high-priority work
immediate access to the CPU when %
runnable, while scheduling low-priority tasks only whenever CPU capacity would be otherwise idle.
To do so, UFS introduces two \emph{scheduling tiers}: a "\high{}"  tier for high-priority work, and a "\low{}" tier for low-priority work. 
Moreover, UFS supports \texttt{cgroups} in each  tier enabling fine-grained priority-based proportional sharing within each tier.

\textit{Priority-inversion.} UFS actively manages the priority-inversion problem. Recent work~\cite{sigmodcomp} indicates that conflicts are no longer a major
issue in current database systems, as they often use optimistic concurrency control, in particular for read-only tasks --- and the data analytics work that we expect to run in the background is purely read-only. 
However, DBMS still use short-lived locks for shared data structures such as the \emph{write-ahead log} (WAL) or certain components within the buffer manager. 
The standard solution to avoid priority inversion is to avoid preempting tasks while such locks are held. We take this a step further: only when a \high\ task waits for a lock held by a \low\ task, do we avoid (or reverse) preemption by temporarily increasing the priority of the \low\ task. We cross the DBMS / OS boundary through what we refer to as \emph{application-based} scheduler hinting: A database task can signal our scheduler whether it is holding or requesting any locks.

\textit{PostgreSQL-based prototype.} Our prototype implementation features PostgreSQL as a popular open-source DBMS that currently relies on the OS to schedule concurrent tasks. This allows us to holistically explore various options within a realistic setting. We use PostgreSQL's extension system to allow users to conveniently declare priorities for workloads. We implement the application-based scheduler hinting directly in PostgreSQL.

\myparagraph{Contributions}
This paper makes the following key contributions:
\begin{new-item}
\item \textbf{Detailed analysis of state-of-the-art OS schedulers.} We evaluate existing OS scheduling algorithms under mixed database workloads. We show how and why Linux scheduling options disadvantage \high{} CPU-bursty transactional workload in the presence of CPU-bound DB analytics.

\item \textbf{A selectively unfair scheduler.} We design the selectively unfair scheduler UFS for mixed workloads. It grants \high\ tasks immediate CPU access, schedules low-priority background work via low-overhead deferred dispatch,  and preserves \texttt{cgroup}-proportional sharing and constraints.

\item \textbf{Minimally invasive prototype.} We present a fully integrated prototype implementation, based on PostgreSQL hooks and Linux's eBPF subsystem with the \texttt{sched\_ext} interface. Our solution requires only minimal code modifications to PostgreSQL and no modifications to the Linux source code, making it easy to adopt and maintain in the long term. 

\item \textbf{An SQL interface to assign process priority.} We provide a PostgreSQL extension that enables users to assign priorities to their PostgreSQL sessions directly in SQL. %

\item \textbf{Application-hinting.} To address the priority inversion problem, we enable a line of communication between DBMS and OS scheduler, by leveraging eBPF maps.

    \item \textbf{Experimental evaluation.}
    We emulate mixed database workloads and provide evidence that UFS can assign CPU-bursty tasks highest priority when run concurrently with background tasks. Ultimately, this improves throughput by up to 2$\times$ while reducing tail latency by half, compared to Linux's proportional sharing scheduler EEVDF. Even Linux real-time scheduling has tail latencies 1.5$\times$ higher than UFS. %
    At the same time, UFS provides proper proportional sharing when CPU-bursty and CPU-bound workloads have equal priority.
    Performance remains as expected with various \texttt{cgroups}, different hardware configurations, and workloads. 
   To conduct controlled experiments with priority inversion,
    we show in a micro-experiment that application-hinting indeed succeeds at avoiding the priority inversion.

\end{new-item}

\myparagraph{Structure} The remainder of this paper is organized as follows. Section~2 outlines techniques used. Section~3 discusses the shortcomings of existing Linux schedulers confronted with mixed database workloads. Section~4
motivates the development of UFS and describes its conceptual design, while Section~5 provides implementation details. Section~6 evaluates UFS. Section~7 discusses the related work, and Section~8 concludes the paper.

\section{Background}

To facilitate understanding of UFS, we provide background on cgroups, Linux scheduling, eBPF, sched\_ext, and PostgreSQL.

\myparagraph{Control Groups}
The \emph{control groups} (cgroups) feature of the Linux kernel can be used to organize processes into hierarchical groups and to apply resource-management policies~\cite{linux_cgroups_docs}. 
Commonly, cgroups are utilized to isolate resources in container runtimes.%

A cgroup can impose limits and accounting for resources (e.g., CPU time, memory, and I/O). For CPU scheduling, a weight parameter (\texttt{cpu.weight}) is exposed that represents the cgroup’s relative share of CPU time: %
Under contention, a cgroup with \texttt{cpu.weight} set to~200 is intended to receive roughly twice as much CPU time as a cgroup with weight~100. 
By placing processes\footnote{Even individual threads or tasks can be assigned to different cgroups.} into different cgroups and dynamically adjusting their weights, Linux can allocate more CPU time to high-priority workloads while still ensuring proportional fairness among all runnable groups.

\myparagraph{Scheduling in Linux}\label{sec:scheduling-classes}
The Linux kernel provides different schedulers. Each scheduler is responsible for handling the tasks of a \emph{scheduling class}. The kernel maintains a strict static precedence order. When the kernel searches for tasks to execute, it first considers runnable tasks in the scheduling class with the highest precedence, then proceeds to lower-precedence classes. Within these classes, a \emph{scheduling policy} specifies the exact rules for scheduling a task. %

The classes discussed in this paper are ranked by precedence from highest to lowest as follows: (1) real-time scheduling with the policies \texttt{\FIFO{}} and \texttt{\RR{}}, (2) fair scheduling with the policies \texttt{SCHED\_NORMAL} and \texttt{SCHED\_IDLE}, and (3) eBPF-based scheduling with the \texttt{SCHED\_EXT} policy to schedule tasks via the extensible \texttt{sched\_ext} scheduler framework. %

\myparagraph{Real-time scheduling}
The %
policies \texttt{\FIFO{}} and \texttt{\RR{}} are used to schedule tasks with real-time requirements, based on static priorities ranging from 1 to 99. \texttt{\FIFO{}} uses a first-in-first-out policy: A runnable task is executed as long as it is runnable, unless it is preempted by a higher-priority real-time task.
In contrast, \texttt{\RR{}} uses round-robin; there is one run queue for each priority. The first runnable task from the highest-priority queue is executed for a time quantum and then moved to the end of the queue, and the next runnable task is executed. Runnable tasks with higher priority preempt tasks with lower priority.

\myparagraph{Fair scheduling} 
Since Linux 6.6, \emph{Earliest Eligible Virtual Deadline First} (EEVDF) has been the default fair scheduler~\cite{linux_sched_eevdf}. EEVDF provides approximate proportional fairness~\cite{stoica1995eevdf} and supports cgroup-based CPU scheduling. It assigns each runnable task that uses the \texttt{SCHED\_NORMAL} policy a virtual deadline based on its weight (higher-weight tasks receive earlier deadlines and more CPU share) and runs the eligible task with the earliest virtual deadline.
A task under the \texttt{SCHED\_IDLE} policy runs only when CPU cycles are available.

\myparagraph{eBPF}
The \emph{extended Berkeley Packet Filter}~\cite{linux_bpf_docs} is a Linux kernel subsystem that enables runtime loading and execution of small programs within the kernel. The programs are \emph{eBPF verified} before usage and compiled just-in-time. %
eBPF programs are triggered by events such as function calls in user space applications or by specific kernel hook points. 
eBPF programs are stateless but can store data in \emph{eBPF maps}, enabling them to maintain state between invocations. 
More complex data structures, like ring buffers and red-black trees, are built on top of these maps. Furthermore, these eBPF maps can be read and written from user space. %

\myparagraph{sched\_ext}\label{sec:sched_ext}
The behavior of the \texttt{sched\_ext} scheduling class is determined by dynamically loaded eBPF programs~\cite{kerneldocs_sched_ext} that implement a specific scheduling policy. This allows scheduler policies to be implemented as user-loadable eBPF programs. \texttt{sched\_ext} provides hook points that are called when scheduling decisions are to be made, such as when a task becomes runnable, or a CPU needs work. The actual scheduling is performed by %
eBPF callback implementations registered on these hooks. Callbacks that are not implemented fall back to default behavior, allowing a scheduler to focus on the parts it needs, while relying on the default implementation for the rest. \texttt{sched\_ext} can operate in partial mode, only scheduling tasks that are assigned to the \texttt{SCHED\_EXT} policy, or in full mode, where it also schedules tasks under \texttt{SCHED\_NORMAL} and \texttt{SCHED\_IDLE}.

A \texttt{sched\_ext} policy implementation also controls where runnable tasks wait before they are executed. 
    Local \emph{dispatch queues} (DSQs) are per-CPU run queues that hold tasks intended to run on a specific CPU soon. Since they are private to one CPU, they are typically small and fast, and they are the first place a CPU checks when selecting its next task. %
    Once a task is placed in a CPU’s local DSQ, it stays until the CPU selects it for execution. When the task starts running, it is removed from the local DSQ. Once execution stops, the scheduler must place it in the DSQ before it can run again.
    In contrast, 
    custom DSQs are policy-specific queues, such as per-cgroup queues, or queues for priority classes. Tasks in custom queues do not execute immediately; they remain queued until a CPU becomes available and the scheduler dispatches them to a CPU-local DSQ.

\myparagraph{PostgreSQL}\label{sec:pg-locks}
PostgreSQL is one of the most popular open-source relational DBMS. It uses a process-based architecture; each client connection is handled in a separate process, called \emph{backend}.
PostgreSQL uses \emph{multi-version concurrency control} (MVCC) and read-only transactions do not set row-level locks. They do set shared locks on tables but only  conflict with transactions that perform structural changes (e.g., deleting a table). Update transactions can set conflicting row-level locks, but we assume that such transactions never run as \low\ tasks. 

PostgreSQL uses \emph{lightweight locks} (LWLocks) to synchronize access to data structures in shared memory (e.g., transaction ID generation) and \emph{spinlocks} for short-term locks held for just a few instructions. Spinlocks could keep the process in a busy-wait loop, so the system ensures that the process performs short sleep cycles, which help to release the CPU for other tasks. If a spinlock cannot be acquired for some time, an error with \texttt{PANIC} severity is reported. In \emph{panic mode},  PostgreSQL terminates all processes and reinitializes.

\section{Shortcomings of Current Linux Schedulers}
\label{sec:motivation}

Modern database systems
must be able to handle mixed workloads. Interactive transactions are often CPU-bursty, requesting short amounts of CPU time followed by communication with the client. They should typically always run with high priority as short response times are crucial. CPU-bound tasks might either be able to run in the background when they are not time-sensitive, or they may actually be part of an interactive analysis, in which case they do need reasonable response times. 
As such, a scheduler needs to ensure that \low{} tasks use the CPU only when \high{} tasks do not need it. At the same time, tasks with the same priority should receive CPU time proportionally. 
The question is whether mechanisms that Linux provides for priority-based CPU scheduling suffice for mixed database workloads. Our analysis shows that this is not the case. 
However, failure modes differ between schedulers.

\myparagraph{Setup}
We analyzed the behavior of the schedulers by executing three workload mixes, as described in Table~\ref{tab:workload_mixes}. 
Workload ``SOLO'' serves as baseline. We had 4 PostgreSQL backend workers with exclusive access to 4 cores\footnote{The detailed software and server configuration is presented in Section~\ref{sec:evaluations}.}$^,$\footnote{We selected CPUs in strides of two (0, 2, 4, 6) to confine the experiment to a single NUMA (\emph{non-uniform memory access}) node
and thereby avoid interference from cache-preserving heuristics that prioritize NUMA locality over optimal
placement.}
serving 4 CPU-bursty TPC-C clients.

\begin{table}[t]
\caption{Database workloads, stand-alone (SOLO) and mixed.}
\label{tab:workload_mixes}

\centering
\footnotesize

\setlength{\aboverulesep}{0.2ex}
\setlength{\belowrulesep}{0.2ex}
\setlength{\abovetopsep}{0.2ex}
\setlength{\belowbottomsep}{0.2ex}
\renewcommand{\arraystretch}{0.9} %

\begin{tabular}{lll}
\toprule
\textbf{Workload} & \textbf{CPU-bursty} & \textbf{CPU-bound} \\
\midrule
SOLO 
& TPC-C
& \multicolumn{1}{c}{--} \\

MIN:MAX 
& TPC-C, high-prio
& TPC-H, low-prio \\

50:50 MIX
& TPC-C, high-prio
& TPC-H, high-prio\\
\bottomrule
\end{tabular}
\end{table}

\begin{table}[t]
\caption{Scheduling high and low-priority tasks.}
\label{tab:impl}
\centering
\footnotesize

\setlength{\aboverulesep}{0.2ex}
\setlength{\belowrulesep}{0.2ex}
\setlength{\abovetopsep}{0.2ex}
\setlength{\belowbottomsep}{0.2ex}
\renewcommand{\arraystretch}{0.9} %

\begin{tabular}{lll}
\toprule
\textbf{Scheduler} & \textbf{High-prio} & \textbf{Low-prio} \\
\midrule
EEVDF 
& NORMAL (weight=10k) & NORMAL (weight = 1) \\

IDLE
& NORMAL (weight=10k)
& IDLE \\

FIFO
& FIFO (priority = 99)
& NORMAL (weight=1) \\

RR
& RR (priority = 99)
& NORMAL (weight=1) \\
\bottomrule
\end{tabular}
\end{table}

In the two other workloads, the 4 TPC-C workers ran concurrently with 4 further backend workers executing CPU-bound tasks.
To emulate a read-only CPU-bound workload, we execute a TPC-H query in a tight loop within a UDF. Rather than using a stand-alone ML workload, we deliberately choose a complex analytics query that competes for PostgreSQL-internal data structures and might trigger priority inversion. %

In the ``MIN:MAX'' workload mix, the CPU-bound tasks are considered to be background tasks only. We therefore assign the CPU-bursty tasks maximum priority, and the CPU-bound tasks minimum priority. 
In the ``50:50'' mix, we assume that the CPU-bound TPC-H queries are nevertheless time-critical, so we assign them the same high priority as the CPU-bursty tasks. To achieve these low/high priority combinations, we set the scheduler configurations as outlined in Table~\ref{tab:impl}, using EEVDF and \texttt{SCHED\_NORMAL} with different weights for high- and low-priority tasks\footnote{Note that 1 is the minimum and 10K the maximum weight possible for \texttt{cgroups}.}; using \texttt{SCHED\_IDLE} for low-priority tasks; or using real-time scheduling \texttt{SCHED\_FIFO} or \texttt{SCHED\_RR}\footnote{Note that 99 is the maximum priority for a task in \texttt{SCHED\_FIFO} or \texttt{SCHED\_RR}.} for high-priority tasks, while using \texttt{SCHED\_NORMAL} for low-priority tasks. Note that IDLE is only relevant for MIN:MAX.

\begin{figure}[t]
    \centering
    \includegraphics[width=0.95\columnwidth]{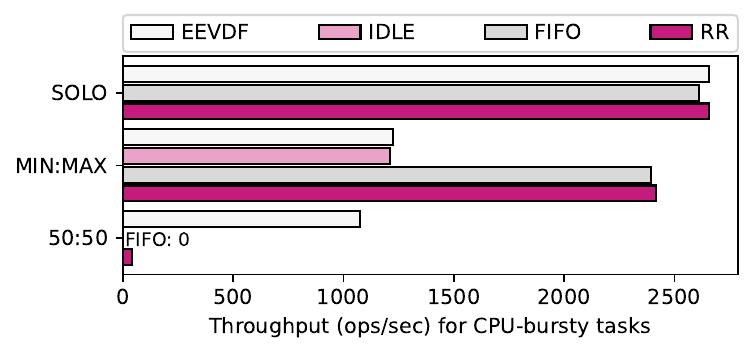}
    \caption{Throughput of CPU-bursty tasks, run stand-alone and as part of mixed workloads. Varying OS schedulers.}
    \label{fig:linux}
\end{figure}

 \myparagraph{Results}
Figure~\ref{fig:linux} shows the throughput for the CPU-bursty tasks with these scheduler configurations. 
SOLO is the baseline, as each CPU-bursty task gets an exclusive CPU. This is the best-case scenario, and all scheduling strategies perform equally. For MIN:MAX, we would expect the CPU-bursty throughput to be nearly the same as SOLO, because CPU-bound tasks have much lower priority. However, with EEVDF and IDLE, throughput suffers considerably.
In contrast, FIFO and RR can still deliver very good throughput. For 50:50, we expect the CPU-bursty throughput to be reduced proportionally, as the CPU-bound tasks should receive an equal share. This can be observed with EEVDF. However, for FIFO and RR, the throughput collapses, even reaching zero in one case.

\myparagraph{EEVDF fails to balance initial task placement}
In EEVDF, when a task wakes up or is forked, it must be
assigned a target CPU before it becomes runnable. EEVDF first chooses a base CPU, typically
favoring the previous CPU or a nearby sibling CPU involved in the wakeup.
When the base CPU is busy, the scheduler scans nearby sibling CPUs in a deterministic order
and returns the first CPU that appears idle.
Since
the background tasks keep most CPUs busy most of the time, CPU-bursty
tasks frequently fail the base-CPU check and fall into the idle-sibling path. %
CPUs that already host CPU-bursty work are more likely to become briefly idle again, so
they are selected repeatedly. As a result, wakeups tend to stack CPU-bursty tasks onto the
same few CPUs instead of spreading them across available CPUs, leading to poor performance for these CPU-bursty tasks.

\begin{wrapfigure}{r}{0.6\linewidth}
    \vspace{-6pt}
    \centering
    \includegraphics[width=5cm]{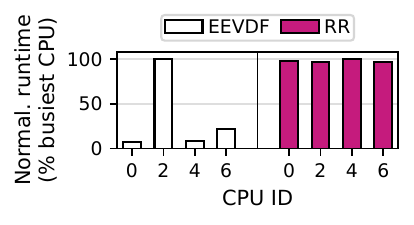}
    \vspace{-6pt}
    \caption{CPU utilization by high-prio workers across 4 CPUs.}
    \label{fig:init_chart_placement_eevdf_rr}
    \vspace{-10pt}
\end{wrapfigure}

Figure~\ref{fig:init_chart_placement_eevdf_rr} illustrates this behavior. We reconstruct the per-CPU execution time for CPU-bursty tasks from kernel scheduler traces (\texttt{sched\_\allowbreak switch} events). For each CPU, whenever a context switch occurred, we computed the elapsed time since the previous switch on that CPU and attributed that interval to the process that had been running. We then sum the total runtime of the processes handling CPU-bursty tasks. %
In Figure~\ref{fig:init_chart_placement_eevdf_rr}, the y-axis shows CPU utilization of these CPU-bursty tasks, normalized to the CPU that has the highest load due to CPU-bursty tasks (100). %
The differences in bar height indicate that some CPUs receive more CPU-bursty tasks than others. In fact, EEVDF frequently concentrates multiple CPU-bursty tasks on the same CPUs, creating short-term pile-ups. The other CPUs continue to be used for \low\ tasks. EEVDF eventually mitigates some pile-ups via periodic load balancing. However, because CPU-bursty tasks are short-running, the skew and imbalance often persists for a large fraction of the request lifetime. By the time load-balancing kicks in, throughput has already been impacted. 
In contrast, RR distributes \high\ work evenly across available CPUs because it can preempt lower-priority work immediately, instead of pushing CPU-bursty work toward CPUs that briefly appear idle.

 \textit{These results show that we need a scheduling strategy with smart initial placement and aggressive preemption. }%

\myparagraph{Real-time scheduling cannot handle same-priority mixed workloads}
While real-time scheduling can handle workloads with different priorities well, it fails when CPU-bursty and CPU-bound tasks have the same priority. Under FIFO, once a task is assigned to a CPU, it runs to termination unless a higher-priority task arrives. Once a CPU-bound task gets the CPU, the CPU-bursty tasks with the same priority are completely delayed. With RR, runnable tasks of the same priority execute in fixed CPU quanta, that is, short bounded turns on the CPU. CPU-bursty interactive tasks frequently block briefly on I/O, locks, or client interaction before consuming their full turn. When that happens, they lose the unused remainder, whereas a CPU-bound worker of the same priority can keep consuming full turns repeatedly. %
The major problem here is that FIFO and RR lack virtual runtime accounting. In contrast, EEVDF calculates virtual runtimes, allowing it to prioritize tasks that have received less CPU time so far, including bursty tasks that repeatedly block for I/O. 

\textit{This shows that our scheduler must support virtual runtime.}

\section{Solution Design}
We introduce the conceptual solution design of our \emph{Unfair Scheduler} (\textbf{UFS}) here and discuss its implementation in Section~\ref{sec:implementation}. 
Our solution aims at DB-specific scheduling that can be implemented without significantly changing the DBMS or the OS. We want to give time-sensitive tasks absolute priority by preempting tasks that should only run in the background when the CPU is available. At the same time, if there are multiple tasks with the same priority, they should be scheduled proportionally. That is, our scheduler needs to avoid the pitfall of the existing OS schedulers.  Furthermore, it must handle the possibility of priority inversion. 

\begin{figure}[bt]
    \centering
    \includegraphics[width=0.9\columnwidth]{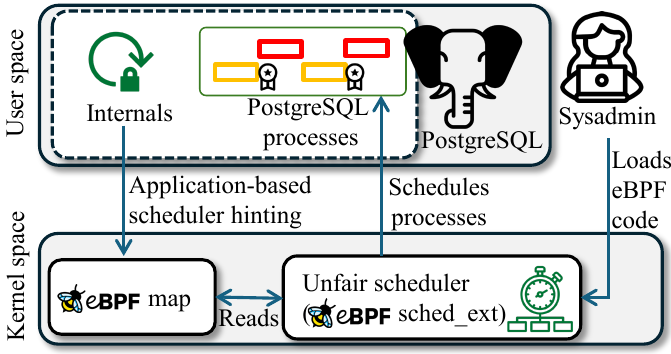}
    \caption{UFS framework architecture.}
    \label{fig:scheduler}
\end{figure}

\textit{Architecture.} Figure~\ref{fig:scheduler} shows the architecture of our solution. The mixed-workload application runs in user space. Here, this is PostgreSQL executing \high\ and \low\ tasks (yellow/red boxes in the figure). 
UFS is a novel scheduler policy implementation built on the \texttt{sched\_ext} class, hence it does not require invasive in-kernel changes.
To install UFS, the system administrator simply loads the eBPF code of UFS at run-time into the Linux kernel space.
The DBMS communicates scheduling-critical internals (such as held locks, see Section~\ref{sec:pg-locks}) to the UFS scheduler by writing to an eBPF map allocated in kernel space.

\myparagraph{Scheduling Tiers, Task Prioritization, and cgroups}
We assign each PostgreSQL backend process, and with it the tasks it executes, to one of two scheduling tiers: (1)~\emph{\high} and (2)~\emph{\low}. Furthermore, in order to manage task priorities among tasks of the same tier, UFS supports Linux cgroups, allowing multiple cgroups to exist within each tier. %
Whether a cgroup belongs to the background or time-sensitive tier is determined by the name of the cgroup.
In return, a task's tier is determined from the cgroup to which it belongs. Task prioritization within a tier is achieved by setting proper CPU weights in the respective cgroups, %
which UFS honors.%

As with weight-based fairness in EEVDF, tasks under UFS inherit the effective share of the cgroup they belong to.\footnote{UFS also honors the standard CPU-controller constraints exposed through cgroups. This includes usage limits (\texttt{cpu.max}) and CPU affinity restrictions (\texttt{cpuset.cpus}), although we do not anticipate that they are relevant for mixed database workloads.} 
As with the conventional cgroup interface, weights and related parameters can be dynamically adjusted. %
UFS respects the cgroup hierarchy so that each cgroup's parameters are defined relative to its parent, with changes propagating %
accordingly. 

To simplify UFS configuration, we provide a PostgreSQL extension that creates and configures the necessary cgroups and assigns the current PostgreSQL backend to the appropriate cgroup based on SQL configuration variables. Assignment can be done dynamically. %

\myparagraph{Task Scheduling}
Tasks are scheduled according to their tier w.r.t.\ (1)~\emph{priority under contention}: which task is allowed to run when CPU time is scarce, and (2)~\emph{CPU selection and dispatch}: how a task is assigned to a CPU and how it reaches that CPU's run queue.

\subparagraph{\em (1) Priority under contention.}
\High\ tasks always take precedence over \low\ tasks, regardless of cgroup weights. This optimizes for mixed database workloads where \high\ corresponds to interactive queries, and \low\ corresponds to batch work, such as analytics queries or machine learning tasks, that are often CPU hungry.
When tasks from both tiers are runnable, UFS prioritizes %
\high\ tasks. UFS schedules \low\ work only when this does not delay \high\ execution.

\begin{figure}[t]
  \centering

  \includegraphics[width=\columnwidth]{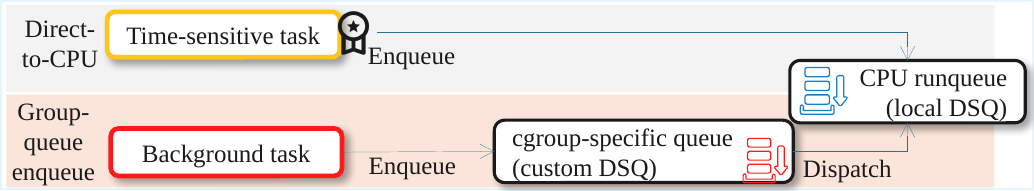}

  \caption{Enqueue strategies.}
  \label{fig:enqueue_strategies}

\end{figure}

\subparagraph{\em (2) CPU selection and dispatch.}
For \high\ tasks, UFS applies the \emph{direct-to-CPU strategy}:
As shown in Figure~\ref{fig:enqueue_strategies}, when a \high\ task becomes runnable, the scheduler selects an appropriate target CPU based on the current system conditions and directly enqueues the task into that CPU's run queue.
The task can run immediately when the target CPU is idle or running background work, in which case UFS issues an idle or preemption kick.
An important aspect of this \emph{direct-to-CPU} strategy is that CPU selection ensures a balanced task placement from the start, rather than relying on later corrective load balancing.

For \low\ tasks, UFS implements the \emph{group-queue enqueue strategy}: When a \low\ task becomes runnable, it is first enqueued into a cgroup-specific queue. Only when a CPU becomes idle and requests work does the scheduler select a task from an eligible \low\ cgroup and dispatch the task to that CPU's run queue. This \emph{deferred} placement has two benefits: (1)~It avoids expensive global CPU searches upon wake-up. (2)~It allows for reactive load distribution by having CPUs pull background work on demand. The scheduler can still respect cgroup proportionality and affinity constraints when deciding which \low\ tasks to dispatch.

\myparagraph{Application-based Scheduler Hinting}\label{sec:application-hinting}
UFS must prevent priority inversion. 
Priority inversion requires at least three concurrent tasks. If a \low{} task $T_1$  holds a lock that a \high{} task $T_2$ wants and blocks, and $T_1$ and $T_2$ are the only two tasks in the system, then $T_1$ will be able to run until it releases the lock and will then be preempted by $T_2$. No priority inversion occurs. However, if there are other \high{} tasks that are runnable, they will preempt~$T_1$, and $T_1$ will not be able to release the lock, delaying $T_2$ with it. $T_2$ involuntarily receives the same low priority as $T_1$. 

Note that PostgreSQL prevents a task $T_2$ from spinning indefinitely when it tries to acquire a spinlock held by another task $T_1$. Instead, PostgreSQL spinlocks spin only briefly before sleeping for a substantially longer interval. This behavior effectively prevents the more direct form of priority inversion in which a time-sensitive task starves the background task that must run to release the lock it is waiting for.
Nevertheless, the indirect inversion described above remains possible, where $T_1$ cannot release the lock because yet another high-priority task occupies the CPU.

To address this, we enable application-based scheduler hinting via an eBPF map: user space applications can write hints into this map, which eBPF programs in kernel space can access.
This enables communication between user space programs and UFS. When a user space application inserts a hint that a \high\ task is waiting for a lock, the scheduler checks whether a conflicting lock is currently held by a \low\ task and temporarily treats that \low\ task as runnable in the \high{} tier until the lock is released. Note that this allows \low\ tasks to be preempted when they hold locks, but only as long as no \high\ tasks want the locks.

\section{Implementation}
\label{sec:implementation}
This section describes our prototypical implementation of UFS. We first outline our implementation of UFS based on \texttt{sched\_ext}, then the small-scale adjustment to the DBMS code for application-based scheduler hinting, and finally the user-facing PostgreSQL extension.

\subsection{Unfair Scheduler}
UFS is built on the \texttt{sched\_ext} framework using callbacks and DSQs (see Section~\ref{sec:sched_ext}). Before we discuss the implementation of the \emph{enqueue} and \emph{dispatch} callbacks, which are the key components of UFS, we introduce the concept of virtual runtime.

\subsubsection{Virtual runtime}
UFS allocates CPU time in \emph{time slices}, i.e., hard-coded bounded execution intervals during which a task or cgroup is allowed to run before scheduling may be reconsidered. Virtual-runtime scheduling in UFS tracks how much CPU time a task has received across such slices, so that tasks that have received less service become more eligible to run. UFS tracks virtual runtime on two levels: (1)~\emph{task virtual runtime} tracks the runtime of tasks \textit{within} a cgroup, scaled according to the cgroup's weight (\emph{weight-scaled virtual runtime}), and (2)~\emph{cgroup virtual runtime} tracks the sum of the virtual runtimes of all tasks in the cgroup.

UFS chooses the cgroup with the \emph{lowest} cgroup virtual runtime as the next one to receive CPU service. Within that cgroup, it then selects the task with the \emph{lowest} task virtual runtime to run next. This two-level mechanism is needed because in \texttt{sched\_ext}, weights are assigned at the level of cgroups rather than tasks: cgroup virtual runtime determines \emph{which} cgroup should receive service next according to its weight, while task virtual runtime determines \emph{which} task within that cgroup should run next in a fair manner. Note that this implementation differs from EEVDF, as EEVDF uses virtual deadlines rather than virtual runtime ordering.

\subsubsection{Enqueue Callback}
The enqueue callback is triggered whenever a task becomes runnable (e.g., upon wake-up) or must be re-enqueued (after preemption, yield, or slice expiration). Our callback implementation involves: (1) looking up the task's state and cgroup, (2) clamping the task's virtual runtime to prevent credit hoarding, and (3) enqueuing the task.

\textit{State lookup:}
UFS first retrieves the task's virtual runtime and identifies its current cgroup, as provided by the kernel. It then loads the corresponding per-cgroup scheduling state, including metadata such as the cgroup's weight, class, and virtual runtime.

\textit{Clamping virtual runtime:}
Each task maintains a virtual runtime, reflecting its accumulated runtime, and used to order tasks within its cgroup. Before inserting the task into a DSQ, the scheduler limits how far behind the cgroup’s current virtual runtime a task may lag. Specifically, if the task’s virtual runtime is older than ``one task slice'' relative to the cgroup’s current virtual runtime, it is raised to that bound. This prevents a task that has been idle for a long time from accumulating scheduling credit and immediately jumping ahead of the cgroup’s recently active tasks. Virtual runtimes of \high\ tasks are inherently kept lower than those of \low\ tasks, so that they are always prioritized on a given local DSQ.

\textit{Enqueuing:}
UFS implements two enqueuing strategies depending on the task tier: (1)~\emph{direct-to-CPU} for \high\ tasks and (2)~\emph{group-queue enqueue} for \low\ tasks (see Figure~\ref{fig:enqueue_strategies}).

\begin{new-item}
    \item \textit{Direct-to-CPU enqueue:} The task is inserted into a CPU-local DSQ, and immediate rescheduling is performed if needed. In detail, UFS determines a CPU that can run the task promptly. The task is then enqueued in the CPU's DSQ. If there are already other \high\ tasks in the queue, its virtual runtime is used to determine the queue position. Otherwise, UFS issues a CPU ``kick'': It wakes the CPU if it is idle or requests preemption if the CPU is running \low\ work.

\item \textit{Group-queue enqueue:} The task is inserted into its cgroup’s custom DSQ using the virtual runtime as key. Since wake-up latency is not a concern, direct CPU placement is intentionally avoided. The dispatch callback is responsible for moving these tasks eventually into a local DSQ.

\end{new-item}

\subsubsection{Dispatch Callback}\label{sec:dispatch}
The dispatch callback is invoked whenever a CPU needs work and its DSQ is empty. It means that no \high\ tasks need the CPU at the moment. Therefore, the dispatch callback must check whether a runnable \low\ task requires CPU and move this task to the CPU-local DSQ. 

\begin{figure}[tb]
  \centering
\includegraphics[width=0.7\columnwidth]{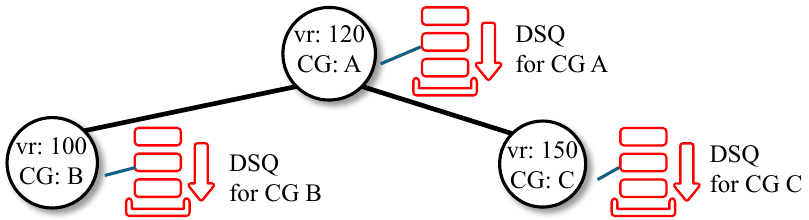}
  \caption{A \runnabletree\ with nodes for cgroups~A, B, C.%
  }
  \label{fig:dispatch}
\end{figure}

\myparagraph{\Runnabletree}
UFS maintains cgroup semantics, in particular preserving scheduling proportional to weight. To achieve this, it uses a data structure that we call the \emph{\runnabletree}. It manages the runnable background cgroup DSQs. Since eBPF is used, the scheduler must use bounded, eBPF verifier-friendly data structures such as maps and DSQs. Therefore, we implement the \runnabletree\ as a binary search tree based on the existing eBPF red-black tree, as shown in Figure~\ref{fig:dispatch}. Each node in this tree represents a cgroup DSQ, using its virtual runtime as the key. The leftmost leaf of the tree indicates the cgroup with the lowest virtual runtime.

When UFS tries to find the next \low\ task to run, it selects the next eligible cgroup DSQ from the \runnabletree\ and transfers the task that has run the least so far into the CPU-local DSQ. This is done as follows:

\myparagraph{Selecting the next cgroup DSQ}
UFS repeatedly tries (up to a small bounded number of iterations) to obtain work as follows: (1) Peek the cgroup with the minimum virtual runtime from the \runnabletree. (2) Verify active state. If the cgroup disappeared (e.g., tasks exited), dispatch removes the %
node from the \runnabletree\ and retries. (3) Try to obtain runnable tasks. Dispatch attempts to move the runnable task with the smallest virtual runtime from the selected cgroup’s queue to the CPU’s local DSQ. There are two outcomes:

\begin{new-item}
    \item \textit{Task found:} the CPU commits to this cgroup as its current cgroup. Its virtual runtime is then advanced by one time slice, scaled inversely by the cgroup's effective weight, and the cgroup is reinserted into the \runnabletree, with the updated virtual runtime as key. This allows sharing proportional to weight across background cgroups without scanning.
    \item \textit{Empty:} if the cgroup has no remaining enqueued tasks, dispatch removes the cgroup from the \runnabletree\ and places the corresponding bookkeeping node into a per-cgroup stash
so it can be reused on the next enqueue. This avoids keeping empty cgroups in the \runnabletree.
\end{new-item}

\myparagraph{Selecting the next task within a cgroup DSQ}
Within each cgroup DSQ, tasks are ordered by their virtual runtime. The task with the lowest virtual runtime has executed the least amount of work so far and is positioned at the front of the queue. When UFS selects that cgroup to run, the task at the head of its custom DSQ is moved to a CPU’s local DSQ and thus runs first.

Weight-scaled virtual runtime enables weight-proportional scheduling: task virtual runtimes, and therefore cgroup virtual runtimes, advance according to their assigned cgroup weights. Higher-weight tasks and cgroups are then selected proportionally more often.

With group-queue enqueue, dispatch implements a slice-based policy proportional to cgroup weight using the \runnabletree\ data structure. Dispatch selects the earliest eligible cgroup, removes truly empty cgroups, and charges cgroups by advancing their virtual runtime before reinserting them for future selection.

\subsection{Application-Based Scheduler Hinting}
\label{sec:sched-hinting}

UFS relies on user space hints, stored in an eBPF map, to prevent priority inversion. With PostgreSQL, we feed lock events and their completion into this eBPF map. Each map entry contains: (1)~the PID of the \high/\low\ worker
interacting with the lock and (2)~a unique identifier for that lock.

PostgreSQL provides generic infrastructure for reporting \emph{wait events}. These occur in situations where a blocking operation is executed, such as waiting for I/O, reading data from a network client, or attempting to acquire a lock. To update the application-based hints, we add our own calls along PostgreSQL's wait-event reporting path so that lock activity is reported when a lock is attempted, acquired, or released. These calls are inserted at the same points where PostgreSQL invokes the methods \texttt{pgstat\_report\_wait\_start()} and \texttt{pgstat\_re{\allowbreak}port\_wait\_end()}. Our changes to PostgreSQL are very lightweight; we require fewer than 200 lines of code.%
\footnote{The direct changes in PostgreSQL are necessary because these wait events are not propagated via hooks to extension code. In a further version of our implementation, we could just add a new hook for reporting these wait events to PostgreSQL extensions and handle the actual hinting entirely in extension code. This would make the changes in PostgreSQL less invasive: adding a new hook definition would be $\approx$20~lines of code.}
On the side of UFS, we need fewer than 100 lines to process these hints.

\subsection{PostgreSQL Management Extension}\label{sec:options}
By default, PostgreSQL backends are in the \low\ scheduling tier in a default cgroup. To change the tier or cgroup from within PostgreSQL, we provide a PostgreSQL extension that exposes two configuration parameters: (1)~one to specify the workload tier (\high\ or \low) and (2)~one to specify the weight of the associated cgroup. 
Should no cgroup for that tier exist with the given weight, such a cgroup %
is created automatically. 

Since regular PostgreSQL \emph{Grand Unified Configuration} (GUC) parameters are exposed by our extension, they can be manipulated directly via SQL. A client can assign a tier (\texttt{SET task\_tier = ...;}) and weight (\texttt{SET task\_weight = ...;}) to change these values for the current session. By using \texttt{SET LOCAL}, these values can be set at transaction scope. Since these values are managed through PostgreSQL's existing configuration mechanisms, role-level defaults can be set; each time this role connects to PostgreSQL, these predefined values are automatically loaded. %

\section{Experimental Evaluation}
\label{sec:evaluations}
We evaluate UFS and compare it with Linux default scheduling under a variety of settings and workloads. We also analyze application-hinting in micro-experiments and evaluate its overhead.

\myparagraph{Implementation} We implemented UFS in~C
on Arch Linux, targeting Linux kernel~6.17. The loader was compiled with GCC~15.2.1, while the scheduler eBPF program was compiled with Clang~22.1.1. Our database extensions are integrated into PostgreSQL~17.0. Our implementation comprises just about 3,250 lines of C~code, including about 250~lines for the loader, 2,750 lines for the eBPF program, and 250 lines for the PostgreSQL management extension.

\myparagraph{Execution environment}
We use servers within the same network: (1)~PostgreSQL with UFS runs on a dual-socket server with two 24-core Intel Xeon Gold 6248R processors and 400GB of DRAM. Each processor has a base frequency of 3.0GHz and 36MB of cache. (2)~The database clients run on a separate dual-socket server with two 24-core Intel Xeon Gold 6242R processors and 200GB of DRAM. Each processor has a base frequency of 3.1GHz and 36MB of cache.

\myparagraph{Workloads}
In most of our tests, we use the same workloads that were already described in Section~\ref{sec:motivation}. As CPU-bursty work, we run BenchBase~\cite{DifallahPCC13}
(commit \texttt{46fc66f}) on the client machine and execute TPC-C. We deviate from  the default BenchBase configuration and choose scale factor~24. We vary the number of clients. Each client is served by a different PostgreSQL backend (worker). As CPU-bound work, we start UDFs in PostgreSQL at the beginning of each benchmark run.
Each UDF issues TPC-H Query~17 in a continuous loop. 
We also include a test with 
  \verb!schbench!~\cite{mason2016schbench} (commit~\texttt{6300b8f}), a standard benchmark for Linux schedulers, a test with an ML task, and a micro-application for testing application-hinting. Details of these applications are provided in the respective subsections.

\myparagraph{Schedulers} We use the same scheduler configurations, EEVDF, IDLE, FIFO, and RR, with corresponding weights/priorities as outlined in Table~\ref{tab:impl} to handle low-priority \low\ tasks and high-priority \high\ tasks. Using our UFS scheduler, low-priority tasks are in a \low{} tier cgroup with weight~1, and high-priority tasks are in the \high\ tier with weight~10K.

\myparagraph{Configuration}
Simultaneous multithreading (SMT) is enabled on CPUs. To reduce run-to-run variability, we disable turbo boost and restrict the CPUs to the C1 C-state (avoiding transition overheads to deeper C-states). Using cores from both NUMA sockets, we capture any cross-node migration overhead caused by poor placement.

We use a fixed set of CPUs (mostly eight) from the server for our workload, assigned either using cgroup functionality or the \texttt{taskset} feature of FIFO/RR. UFS does not operate in partial mode, so it schedules all tasks of the OS, but we do ensure that our workload has exclusive access to the designated CPUs.

A 1-minute warm-up phase is followed by 1 minute of measurement. Throughput and latency measurements for CPU-bursty tasks are collected on the client side, directly from BenchBase, the remaining tasks are measured at the server side.

\subsection{Mixed Workload Throughput}
\label{sec:eval-throughput}

We explore whether UFS can appropriately handle the workload varieties for which the original schedulers fail. For that, we repeat the experiment in Section~\ref{sec:motivation} and present it in more detail. Our focus is first on throughput.

\myparagraph{Setup} \label{baseline-scenario}
We evaluate the workload mixes from Table~\ref{tab:workload_mixes}, using 8 cores: SOLO with only 8 CPU-bursty workers, MIN:MAX with 8 \high{} CPU-bursty and 8 \low{} CPU-bound workers, and 50:50 where both the 8 CPU-bursty and the 8 CPU-bound workers have the same (high) priority. We include a second SOLO run with only 8 CPU-bound workers to observe them in isolation.

\begin{figure}[t]
    \centering
    \includegraphics[width=0.95\linewidth]{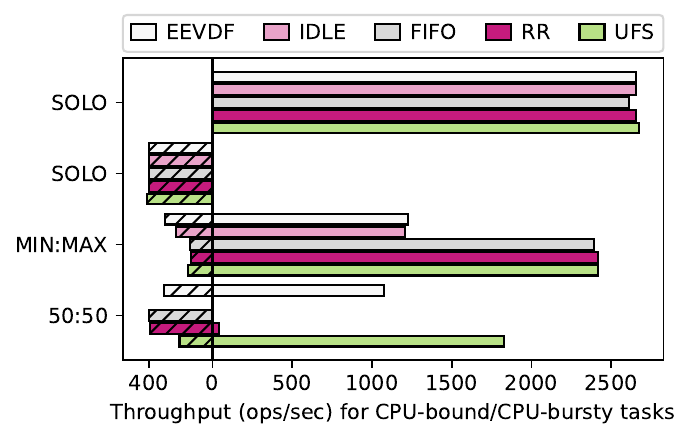}
    \caption{Throughput for CPU-bound vs.\ CPU-bursty tasks (shown left and right resp.), running solo or mixed.}
    \label{fig:solo_vs_mixed}
\end{figure}

\myparagraph{Results}
Figure~\ref{fig:solo_vs_mixed} shows the achieved throughput for \low\ workload (to the left) and for \high\ workload (to the right).
Running these workloads stand-alone (``solo'') serves as baseline against which to compare mixed workloads. In this case, very similar throughput is achieved by all schedulers, both for CPU-bound and for CPU-bursty tasks. This is not surprising, as each PostgreSQL backend can get its own CPU, so scheduling is quite trivial.

For EEVDF, IDLE, FIFO, and RR, we observe the same behavior as in Section~\ref{sec:motivation}. In MIN:MAX, EEVDF and IDLE fail to give the CPU-bursty workloads the requested high priority, reducing their throughput by 50\% compared to running SOLO. Instead, they allow \low{} CPU-bound tasks to reach unexpectedly high throughput. In contrast, FIFO and RR, as well as UFS, are able to keep the throughput of the CPU-bursty tasks high at the expense of the \low{} tasks, offering 2$\times$ the throughput of EEVDF and IDLE.

In 50:50, EEVDF provides the CPU-bursty tasks with reasonable CPU resources, although the share does not appear quite 50:50. The throughput of CPU-bursty tasks is only 40\% of what it is in SOLO, while it is more than 75\% for CPU-bound tasks. With FIFO and RR, the throughput of the CPU-bursty tasks collapses, yet it remains close to 100\% for the CPU-bound tasks compared to running SOLO.
In contrast, UFS provides reasonable performance for both task types: It maintains 75\% of the CPU-bursty throughput and 50\% of the CPU-bound throughput compared to running SOLO.

\myparagraph{Discussion}
The main reasons for the bad performance of EEVDF and IDLE for workload mix MIN:MAX, and FIFO and RR for 50:50 were presented in Section~\ref{sec:motivation}. UFS explicitly avoids the pitfalls.

In the MIN:MAX scenario, in contrast to EEVDF and IDLE, UFS does account for weights and thus, avoids CPU-bursty tasks piling up on the same CPU. Closer analysis confirms that the utilization due to CPU-bursty tasks is the same on all CPUs because it preempts \low{} tasks when \high{} tasks arrive, while for EEVDF, as shown in Section~\ref{sec:motivation}, CPU-utilization varies widely.

For the 50:50 scenario, and as discussed in Section~\ref{sec:motivation}, the bad behavior of FIFO and RR for CPU-bursty tasks is due to the lack of virtual runtime management. Instead, CPU-bursty tasks give up their slice of CPU whenever they block and are then starved by CPU-bound tasks. 
In contrast, as UFS uses virtual runtime accounting, it can provide CPU-bursty tasks a fair share of CPU resources even in the presence of CPU-bound tasks. In fact, it is even ``fairer'' than EEVDF as it ensures that both CPU-bound and CPU-bursty tasks keep at least 50\% of the throughput of the SOLO experiment. In contrast, for EEVDF the initial placement problem still causes a certain amount of imbalance that UFS avoids. 

In summary, UFS's %
aggressive placement for tasks in the \high{} tier and its virtual runtime management combines the respective strengths of Linux's real-time and fair scheduling classes, allowing it to handle challenging database workloads.

\subsection{Mixed Workload Latency}\label{sec:eval-latency-time}

So far, we have only looked at throughput. But for \high\ workloads, it is important to control end-to-end latencies, in particular for CPU-bursty workloads that consist mainly of short transactions. In particular, the long tail of latencies should not contain extreme outliers. We therefore examine the latencies of the \high\ workloads within a mixed workload. Our expectation is that in mixed workloads, UFS produces overall lower latencies for \high\ workloads than EEVDF and RT.

\myparagraph{Setup}
The experimental setup is the same as in Section~\ref{sec:eval-throughput}.

\myparagraph{Results}
Table~\ref{tab:latencies} reports mean and 95th percentile latency summaries for the CPU-bursty tasks for the SOLO, MIN:MAX, and 50:50 scenarios. We omit results for the IDLE and FIFO scheduling options as they were similar or worse than EEVDF resp.\ RR. %

\begin{table}[t]
    \caption{Latencies for CPU-bursty tasks achieved by different schedulers. Highlighting minimum values in bold.}
    \label{tab:latencies}

    \footnotesize 
    \centering

\setlength{\aboverulesep}{0.2ex}
\setlength{\belowrulesep}{0.2ex}
\setlength{\abovetopsep}{0.2ex}
\setlength{\belowbottomsep}{0.2ex}
\renewcommand{\arraystretch}{0.9} %

\begin{tabular}{llrr}
    \toprule
    \textbf{Scenario} & \textbf{Scheduler} & \textbf{Mean latency (ms)} & \textbf{95th percentile latency (ms)} \\
    \midrule
    SOLO   & EEVDF & 3.06 & 5.80 \\
           & RR    & 3.09 & 5.92 \\
           & UFS   & \bf 3.03 & \bf 5.79 \\
    \midrule
    MIN:MAX & EEVDF & 6.63 & 14.23 \\
            & RR    & \bf 3.36 & 9.66 \\
            & UFS   & \bf 3.36 & \bf 6.37 \\
    \midrule
    50:50  & EEVDF & 7.53 & 17.19 \\
           & RR    & 183.24 & 1397.97 \\
           & UFS   & \bf 4.43 & \bf 11.42 \\
    \bottomrule
\end{tabular}

\end{table}

While response times are very similar for the SOLO scenario, for MIN:MAX, EEVDF’s mean latency is 2$\times$ higher than UFS, and its 95th-percentile latency is 2.2$\times$ higher. RR's performance is much steadier than EEVDF, with a mean latency matching UFS. However, its 95th percentile latency remains 1.5$\times$ higher than~UFS. In fact, UFS latencies are only minimally higher than when the CPU-bursty tasks run in isolation. 
For the 50:50 workload, EEVDF's mean latency is 1.7$\times$ and the 95th percentile 1.5$\times$ higher than those of UFS. RR's latencies have completely deteriorated.

\myparagraph{Discussion}
UFS’s advantage extends beyond mean latency: under mixed-priority workloads (MIN:MAX), its aggressive preemption not only largely outperforms EEVDF's response time, but also results in much lower tail latencies, even compared to RR. It can achieve latencies very close to CPU-bursty tasks run in isolation.

For 50:50, UFS achieves a fair and proportional assignment of CPU time to CPU-bursty tasks, despite their bursty behavior. When they are ready to execute, UFS gives them their fair share according to the virtual runtime ordering, providing fairer results than using deadlines as in EEVDF, leading to overall lower latencies.  

\subsection{Throughput Scaling Under Oversubscription}
\label{sec:eval-throughput-wrt-client}

\begin{figure}[t]
    \centering
    
    \includegraphics[width=0.95\linewidth]{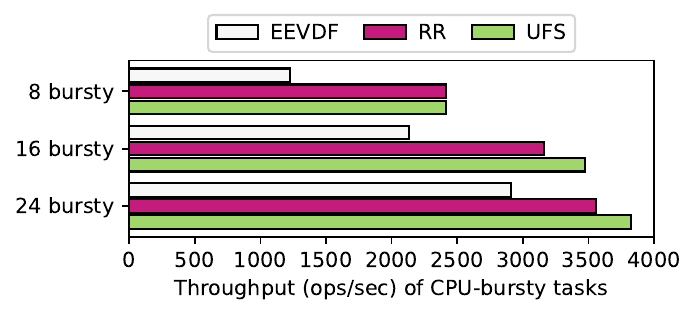}
    \caption{Scaling throughput for CPU-bursty tasks for 8/16/24 CPU-bursty versus 8 CPU-bound tasks.}
    \label{fig:benchmark_clients_scaling_figure_new}
    
\end{figure}

In real database deployments, the number of \high\ CPU-bursty tasks may exceed the number of available CPUs.
We explore how the throughput of \high\ tasks scales under oversubscription in a mixed workload setting.

\myparagraph{Setup}
Our baseline is the MIN:MAX mix from Section~\ref{sec:eval-throughput}, which prioritizes \high\ maximally.
We vary the setup: The \low\ workload remains at 8 workers, the number of \high\ CPU-bursty workers increases from~8, to~16, and then~24.

\myparagraph{Results}
Figure~\ref{fig:benchmark_clients_scaling_figure_new} compares how throughput for the \high\ workload scales under varying degrees of oversubscription,  for different schedulers. As expected, overall throughput increases with increasing number of tasks but not linearly, because \high\ tasks start to compete among themselves for CPU resources. 

Throughput with EEVDF remains well below that of RR and UFS, as observed in Section~\ref{sec:eval-throughput}, but its disadvantage is not as pronounced when there are more \high\ tasks. With 8 \high\ tasks, it achieves only 50\% of the throughput of RR and UFS, compared to 57\% with 16 tasks and 76\% with 24 tasks relative to UFS.  

With increasing workload, UFS starts to outperform RR by up to 10\% when there are 24 \high\ workers. Comparing RR and UFS, UFS performs better. For the baseline, both UFS and RR achieve a throughput 2$\times$ higher than EEVDF. However, as the number of \high\ workers increases, the throughput of UFS exceeds the throughput of RR by almost~10\%.

\myparagraph{Discussion}
The cause for EEVDF's worse performance is its initial placement policy. However, the more \high\ tasks are in the system, the more likely \low\ tasks are already preempted when a \high\ task is scheduled, leading to better placement options and improving EEVDF's relative performance. Still, UFS's aggressive preemption gives it an edge over the others.

\subsection{Mixed Workloads with Varying Priorities}\label{sec:mixed-priorities-eval}

We evaluate how well UFS performs under weight-proportional scheduling, that is, we analyze the performance when it has to  support tasks of different weights in each of the tiers.

\myparagraph{Setup}
We run 16 CPU-bound \low{} workers and 16 CPU-bursty \high{} workers concurrently on the same set of 8 CPUs. The \low{} tasks are equally split into two cgroups with weights 2 and 3, respectively. Similarly, the \high{} tasks are split into two cgroups with weights 6.67k and 10k, respectively. In both tiers, the lower-weight workload therefore has 2/3 of the weight of the higher-weight workload. 

\begin{figure}[t]
    \centering
    
    \includegraphics[width=0.95\linewidth]{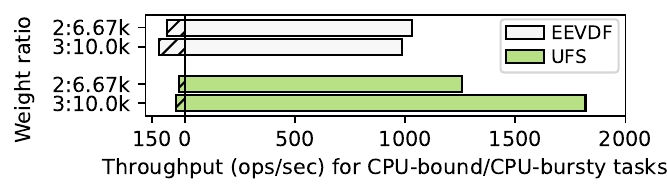}
    \caption{Throughput under different priorities for CPU-bound and CPU-bursty tasks (shown left and right resp.).}
    \label{fig:plot_mixed_priorities}
\end{figure}

\myparagraph{Results}
Figure~\ref{fig:plot_mixed_priorities} shows the %
throughput for each of the four workloads %
for EEVDF and UFS. As seen before, UFS is much better at providing \high\ tasks higher throughput than EEVDF (at the cost of \low{} tasks). 

What is interesting is the relative throughput each workload achieves.
Both UFS and EEVDF preserve proportional CPU allocation among the two \low{} workloads.  With both schedulers, the workload with weight 2 achieves roughly two-thirds of the throughput of the workload with weight 3.

For \high{} workloads, however, the schedulers differ. UFS preserves the expected ratio: \high{} workload with weight 6.67k achieves ca.\ two-thirds of the throughput of \high{} workload with weight 10k. EEVDF, in contrast, gives the two \high{} workloads similar throughput, despite different weights.

\myparagraph{Discussion}
 UFS preserves weight-pro\-portional scheduling not only among background workloads, but also among time-sensitive workloads. This is important as \high{} tasks should preempt \low{} work, but they should be treated according to their weights when they compete with each other.

EEVDF fails %
for the same reason it fails with MIN:MAX in Figure~\ref{fig:solo_vs_mixed}: its wakeup-time placement does not account for weight. Thus, \high{} tasks with different weights are placed in a way that gives them similar CPU time. Later load rebalancing reacts too slowly to correct this for these CPU-bursty tasks.

\subsection{Comparison Under General Workloads}

\begin{figure}[t]

    \centering
    \includegraphics[width=0.95\linewidth]{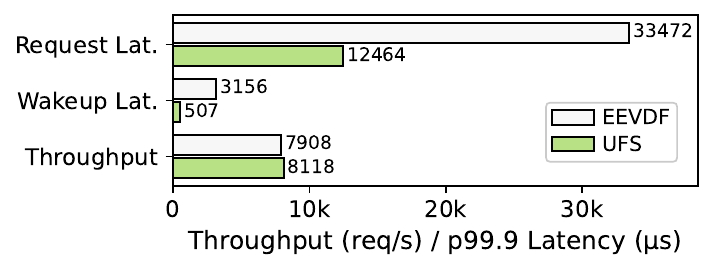}
    \caption{Benchmarking schedulers under general workloads, indicating UFS competes well with the standard scheduler.}
    \label{fig:schbench}

\end{figure}

UFS has been designed primarily for mixed database workloads. We also explore how it performs under the general workloads under which standard OS schedulers are assessed.

We evaluate UFS using 
 \verb!schbench!~\cite{mason2016schbench} (commit~\texttt{6300b8f}),
 a standard benchmark for Linux schedulers.\footnote{Performing 10~sec calibration run, and measuring for 90-second regular run. We use the benchmark's default/example configuration of a 256\,KB cache footprint (\texttt{-F 256}) and 5 operations per compute phase (\texttt{-n 5}), representing moderate settings that preserve some cache pressure and frequent sleep/wakeup behavior.} %
 UFS treats all tasks as ``\low'' and schedules them with default cgroup weight~100.

\myparagraph{Results}
Figure~\ref{fig:schbench} compares the standard scheduler EEVDF with UFS regarding throughput and extreme latency values (reporting 99.9th percentile).
UFS achieves comparable throughput, and features better wakeup and request latencies: UFS reduces wakeup p99.9 latency by 6.2$\times$,
and request p99.9 latency by 2.6$\times$.
    
\myparagraph{Discussion}
These results indicate that UFS is also promising for non-DB workloads. We attribute the lower extreme latencies to the UFS mechanism of pulling background work whenever a CPU is free.
This motivates a more extensive evaluation in future work to assess robustness in practice.

\subsection{Lock-Induced Priority-Inversion}\label{sec:eval-hinting}

As mentioned in Section~\ref{sec:application-hinting}, contention on locks such as spin locks and lightweight locks can induce priority inversion. If a \low{} task holds a lock required by a \high{} task, another \high{} task may starve the \low{} lock holder of CPU time, delaying lock release.
This creates an indirect form of priority inversion in which one \high{} task effectively blocks another through the delayed execution of a \low{} task. This experiment evaluates how effectively UFS can mitigate this effect through our scheduler hinting mechanism.

Under our standard mixed TPC-C and TPC-H workload we were unable to reproduce priority inversion for a sustained period. Although situations did arise in which a background task held a lock needed by a time-sensitive task, the lock was never held for long, since the amount of work performed by the background task while holding it was minimal. Therefore, we created a micro-experiment to illustrate how UFS works to avoid priority inversion.

\myparagraph{Setup}
We implement a PostgreSQL extension used by the two tasks involved in the lock dependency. First, we launch a \low{} task, referred to as the \emph{holder}, which invokes this extension to acquire a designated spinlock, perform a fixed amount of computation (simple arithmetic operations repeated~$10^9 \times$), and then release the lock.  Immediately after starting the \low{} task, we launch a \high{} task, called \emph{waiter}, which invokes the same extension and attempts to acquire the same spinlock; it cannot proceed until the holder releases the lock. We then launch a second \high{} task, referred to as \emph{burner}, that continuously performs CPU-bound synthetic work in a tight loop. All three tasks are pinned to CPU~0. The goal is for the burner to starve the holder of CPU time, thereby indirectly preventing the waiter from making progress.

We compare EEVDF, FIFO, RR, and UFS as configured in the previous experiments. 
Additionally, we have a baseline for comparison, where we run the experiment with only the \low{} holder and the \high{} waiter, and without burner. In this case, the holder will  execute until it releases the lock. This baseline captures the time it takes for the holder to acquire the lock, perform its computation, and release the lock without interference from a concurrent task. When a scheduler is able to avoid inversion, time-sensitive waiters should receive the lock in a reasonable time. %

\begin{table}[t]
    \setlength{\tabcolsep}{3pt}
    \caption{Acquisition and total execution times for the holder and the waiter, reported in minutes and seconds. }%

    \label{tab:spinlock-benchmark}

\centering
\footnotesize 

\setlength{\aboverulesep}{0.2ex}
\setlength{\belowrulesep}{0.2ex}
\setlength{\abovetopsep}{0.2ex}
\setlength{\belowbottomsep}{0.2ex}
\renewcommand{\arraystretch}{0.9} %

    \begin{tabular}{lcccc}
        \toprule
        \textbf{Sched.} & \textbf{Holder acq.} & \textbf{Holder tot.} & \textbf{Waiter acq.} & \textbf{Waiter tot.} \\
        \midrule
        Baseline & 00:00 & 00:03 & 00:04 & 00:04 \\
        EEVDF & -- & -- & -- & -- \\
        FIFO  & 00:00 & 01:11 & -- & -- \\
        RR    & 00:00 & 01:11 & 01:11 & 01:11 \\
        UFS   & 00:00 & 00:07 & 00:07 & 00:07 \\
        \bottomrule
    \end{tabular}

\end{table}

\myparagraph{Results}
Table~\ref{tab:spinlock-benchmark} reports how long the holder and waiter take to acquire the spinlock and to complete. A dash indicates that the experiment did not run to completion.
Compared with the 4-second baseline, adding the burner exposes severe priority inversion under the existing Linux scheduling policies.

Under EEVDF, the experiment fails to complete because PostgreSQL emits a \texttt{PANIC}-severity log message and restarts (see Section~\ref{sec:pg-locks}). This log message is emitted when %
the system is stalled after 1000 consecutive failed attempts to acquire a spinlock.

RR, on the other hand, manages to complete the experiment before PostgreSQL panics, although the waiter still takes more than a minute to acquire the lock. This is due to the fair server, a Linux kernel feature that ensures tasks under the default policy (\texttt{SCHED\_NORMAL}) receive at least 5\% of the runtime on a given CPU, preventing real-time tasks from starving normal tasks~\cite{linux_rt_group_sched,bristot2023dlserver}.

FIFO shows a similar delay in releasing the lock, with the holder again requiring more than a minute to complete its computation. However, unlike RR, FIFO does not use time slices among runnable tasks of the same priority. As a result, once the lock is released, the burner continues to monopolize the CPU, preventing the waiter from running and causing the experiment to stall.

UFS, through application-based scheduler hinting, completes the experiment in only 7 seconds, which is roughly twice the baseline. The holder task is boosted and temporarily treated as a \high{} task, allowing it to receive half of the runtime on CPU~0, which it shares with the burner task.

\myparagraph{Discussion}
While we were not able to measure the effect of any form of priority inversion with our standard mixed workloads, it can occur, in particular when users write their own UDFs with complex computational logic that use locks without much care. We show that UFS has an effective mechanism to avoid such inversions -- something standard Linux schedulers cannot provide, as they have no awareness of application semantics.

\subsection{Overhead of Application Hinting}

Application-level hinting requires PostgreSQL  to exchange information with the OS-based scheduler via eBPF maps, raising the question of potential overhead.

\myparagraph{Setup}
We ran UFS with the MIN:MAX configuration as described in Section~\ref{sec:eval-throughput} with application-hinting enabled and disabled.

\myparagraph{Results}
Throughput differences were negligible ($\leq 1$\%) for runs with and without application-based hinting and within the range expected from run-to-run variability.

\myparagraph{Discussion}
These results indicate that application-based scheduler hinting can act as a low-overhead mitigation mechanism for priority inversion: it is available when lock-induced priority inversion occurs, yet adds no measurable overhead in regular mixed workloads where such inversions do not impact performance. This follows from the design of our PostgreSQL instrumentation, described in Section~\ref{sec:sched-hinting}: hint-related work is triggered only in the wait-event paths, that is, when \high{} tasks wait for locks, avoiding interference with workers when there are no conflicts.

\subsection{ML-based Workloads}
UFS was designed not only to support complex analytical SQL, but also lengthy ML jobs that, if run within the DBMS, should not interfere with interactive workloads. 
As such, this section evaluates UFS with an in-database machine-learning background workload.

\myparagraph{Setup}
We reuse the workloads and configurations from Section~\ref{sec:eval-throughput}, but this time the CPU-bound background workload is ML-based.
The ML tasks run inside PostgreSQL using Apache MADlib\footnote{In order to run Apache MADlib in PostgreSQL~17, we built from commit \texttt{d03af81} with the PostgreSQL~17 support patch from PR~633.}, more specifically, its logistic-regression training routine \texttt{logregr\_train}. Each ML task repeatedly trains a logistic-regression model over a fixed input table, so the workload exercises PostgreSQL execution as well as MADlib's ML implementation.%

\begin{figure}[t]
    \centering
    \includegraphics[width=0.95\linewidth]{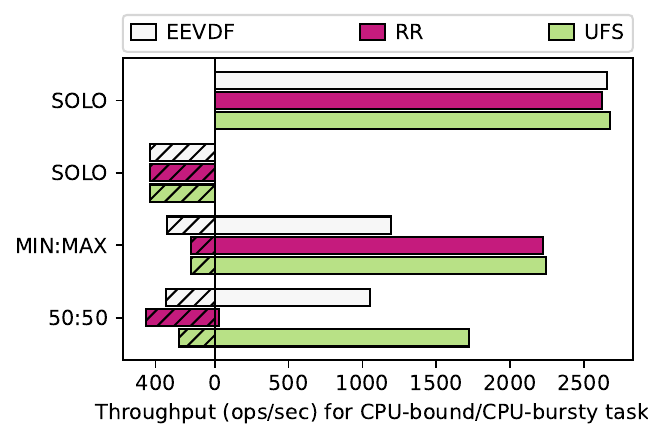}
    \caption{Throughput for a workload mix of CPU-bursty and ML workloads. Left shows MADlib logistic-regression training; right shows CPU-bursty workload.}
    \label{fig:plot_madlib}
\end{figure}

\myparagraph{Results}
Figure~\ref{fig:plot_madlib} shows the throughput results (measured in number of iterations for the regression task). UFS preserves the same qualitative behavior observed for the earlier mixed workloads. In the MIN:MAX configuration, UFS outperforms EEVDF by better protecting the CPU-bursty workload from the background ML tasks. In the 50:50 configuration, UFS again outperforms RR while maintaining a fairer balance between the workloads than EEVDF.

\myparagraph{Discussion}
These results indicate that UFS's benefits are not limited to traditional mixed database workloads. Its main feature is that it can handle mixes of both CPU-bursty and CPU-bound work regardless of the actual computation. As such, UFS is a promising approach for emerging in-database ML. %

\section{Related Work}

Scheduling has been well explored both by the database and the operating system communities.

\subsection{Database Scheduling}

\myparagraph{Concurrency control\!} %
focuses on isolating conflicting concurrent transactions. %
Many database systems are now multi-version and use optimistic concurrency control for read-only transactions~\cite{Cheng24,Deuteronomy,Cicada,speedy}, avoiding locks for reading.
But how runnable transactions access the CPU is  traditionally left to the OS. %

\myparagraph{Resource Scheduling}
Assigning priorities to different database workloads and scheduling their execution accordingly has received attention throughout the years~\cite{CareyJL89,McWherter05,polaris,sigmodcomp}. %
PreemptDB~\cite{sigmodcomp} is a recent proposal and has some similarity to our solution. Transactions can be declared as either low- or high-priority, and low-priority tasks are preempted when high-priority transactions arrive. PreemptDB uses userspace interrupts in order to enforce preemption within the OS kernel. Code snippets that might lead to priority inversion must be declared as non-preemptible regions. 
Different to UFS, %
PreemptDB executes a high-priority transaction entirely before scheduling the next transaction on the same CPU. As such, being implemented at the database layer, it cannot properly handle blocking events such as I/O. In contrast, UFS gives the CPU to another task should one task block, may run several high-priority tasks concurrently on the same CPU, or migrate them when beneficial. It also supports differently weighted tasks within the same workload tier. Finally, UFS avoids preemption only when locks held by background tasks are requested by time-sensitive tasks, while PreemptDB is more conservative, never preempting a transaction within a non-preemptible region even if there are no conflicts.

Priorities and the inversion problem have also been discussed in the context of real-time database systems, where  the scheduler must ensure that transaction deadlines are met~\cite{realtimecc,realtimeinversion,57058}. %

Scheduling mixed OLTP/OLAP workloads has been a major aspect in HTAP Systems (hybrid transaction and analytical databases) \cite{SingleStore,HanaHyper,HTAPSurvey,Ermia} where short update-intensive transactions have very different requirements than long-running analytical queries. However, most of the effort has been on concurrency control, efficient database architectures, and data structures~\cite{HTAPSurvey}. %

Finally, recent solutions propose fine-grained CPU scheduling for analytical queries that are split into small morsels, where parallel execution is controlled by the DBMS~\cite{morsel14,morsel19,morsel21,Numa-data}. Bundling transactions into co-routines to fully utilize the CPU has been discussed in~\cite{cotransaction,cotransaction2}. In contrast, UFS is a more general-purpose OS scheduler that is made application-aware. 

\myparagraph{Scheduling outside of the DBMS}
External query schedulers sit on top of the DBMS, avoiding deep modifications of the DBMS~\cite{4497574,10.14778/3749646.3749686}. Their focus is on predicting query execution time for concurrently executing queries. %
In contrast, UFS resides in the OS below the DBMS and deals with any kind of prioritized queries. %

\myparagraph{Resource Management in Enterprise DBMS}
Commercial systems like Oracle or DB2 allow the specification of service classes with different priorities that are associated with resource constraints (such as CPU shares). The mechanisms might be implemented within the DBMS~\cite{ibm_db2_wlm_redbook_2008,oracle_resource_manager_19c} or the OS~\cite{ibm_db2_13_zos_performance_topics}, but the available documentation does not contain implementation details. It is also not clear whether priority inversion can be handled.

\myparagraph{DBMS within the Operating System}
Offloading DBMS tasks to the operating system and bypassing the user space via eBPF is increasingly being explored. For example, %
\emph{Tigger}~\cite{Tigger} is a connection proxy for PostgreSQL that uses eBPF to forward network packets directly within the kernel, avoiding context switches and data-copy overhead.
\emph{BPF-DB}~\cite{BPF-DB} is a kernel-embedded DBMS that offers ACID-compliant transactions to eBPF applications. 
The authors of~\cite{DB-OS1,DB-OS2} propose to fully rethink how a DBMS could run with and/or in connection with a specialized OS.
In contrast, our goal is to neither change the DBMS nor build a completely new OS. Instead, we build a bridge between DBMS and OS by using eBPF  to develop a DBMS-specific priority-based scheduler.

\subsection{Operating System Scheduling}

\emph{Priority-based Scheduling} has long been supported by operating systems. Unix has included the \texttt{nice} utility since 1973 \cite{ThompsonRitchie1973UnixV4}. The fair scheduler CFS and then EEVDF and cgroups followed. %

There exists previous work that analyzes shortcomings of these priority-based ``fair'' schedulers, including the negative impact of low-priority CPU-heavy analytics on high-priority interactive workloads~\cite{DBLP:conf/eurosys/LeverichK14,DBLP:conf/eurosys/LoziLFGQF16}. However, the accompanying patches are explicitly presented as workload-specific and ``not intended as generic bug fixes'' due to potential side effects~\cite{LoziWastedCoresPatchesNote, ZijlstraLKMLWastedCores2016}.
Some work in the open-source community provided similar evidence of poor performance and also offered patches~\cite{chen2023sis_current,desnoyers2023wakeup_bias_prev_idle,chen2023reply_wakeup_bias_prev_idle}. 
Several of them hinted at the same underlying concern of CFS's and EEVDF's initial idle-CPU placement and wake-up heuristics behaving poorly with short-lived tasks.
Ultimately, none of the proposed patches has ever been merged upstream. This is understandable, %
as maintainers are cautious about workload-sensitive heuristics that may improve a narrow class of workloads while risking regressions elsewhere.
Yet with eBPF and \texttt{sched\_ext}, we can implement an application-specific scheduler that runs within an unchanged Linux kernel. Our particular implementation is different from previous patches as we are defining two different tiers and then additionally offer proportional scheduling within each tier.

\myparagraph{Specialized Operating Systems} In the recent past, completely new operating systems and/or schedulers have been proposed, in particular for cloud-based systems where various applications share the same infrastructure and extremely short tasks must co-exist with longer batch tasks~\cite{Kaffes19,Caladan,LibPreemptible,McClureOSR22}. However, in most of these approaches, different applications are assigned different cores in principle, which might then have to be later reassigned dynamically.

\section{Conclusion}

We address co-scheduling of background and time-sensitive workloads on the same hardware by introducing UFS, a selectively unfair, two-tier scheduler that sits at the intersection of DBMS and OS. Implemented as an eBPF program in the Linux \texttt{sched\_ext} framework, UFS enables holistic scheduling with minimal overhead and allows for information exchange between DBMS and OS via eBPF maps in order to mitigate priority inversion. Background tasks run only on idle CPU cycles and are preempted by time-sensitive workloads, while fair, weight-proportional scheduling is maintained within each tier.  A PostgreSQL prototype requires only $\sim$200 lines of code changes. Experiments show that under mixed workloads, UFS improves interactive throughput by up to 2$\times$.

As future work, we plan to extend UFS to incorporate GPU scheduling, in particular for ML-based analytics. The goal will be to leverage GPUs to offload computation, reduce CPU contention, and improve overall throughput for both analytics and OLTP.

\begin{acks}
This project/research was partly funded by the Passau International Centre for Advanced Interdisciplinary Studies (PICAIS) of the University of Passau, Germany.
\end{acks}

\bibliographystyle{ACM-Reference-Format}
\bibliography{newbib,sample-base}

\end{document}